\documentclass[10pt]{article}
\usepackage{graphicx}

\begin{document}

\title{Sparse Approximation of \\
Computational Time Reversal Imaging }

\author{M. Andrecut}

\date{ }

\maketitle

{\par\centering IBI, University of Calgary \par}
{\par\centering 2500 University Drive NW, Calgary \par}
{\par\centering Alberta, T2N 1N4, Canada \par}

\noindent

\begin{abstract}
Computational time reversal imaging can be used to locate the position of
multiple scatterers in a known background medium. Here, we discuss a sparse
approximation method for computational time-reversal imaging. The method is
formulated entirely in the frequency domain, and besides imaging it can also 
be used for denoising, and to determine the magnitude of the scattering
coefficients in the presence of moderate noise levels.
\end{abstract}

\bigskip 


43.60.+d Acoustic signal processing  

43.60.Pt Signal processing techniques for acoustic inverse problems  

43.60.Tj Wave front reconstruction, acoustic time-reversal 

\pagebreak

\section{Introduction}

Acoustic, elastic or electro-magnetic waves scattered by the inhomogeneities
in the medium carry significant information, which can be used to obtain
true images of the investigated domain [1]. Recently, it was shown that
scattered acoustic waves can be time-reversed and focused onto their
original source location through arbitrary media, using a so-called
time-reversal mirror [2]. This important result shows that one can use
computational time reversal imaging to identify the location of multiple
point scatterers (targets) in a known background medium [3]. In this case, a
back-propagated signal is computed, rather than implemented in the real
medium, and its peaks indicate the existence of possible scattering targets.
The current algorithms for computational time reversal imaging are based on
the null subspace projection operator, obtained through the singular value
decomposition of the frequency response matrix [4-9]. Here, we discuss a
different approach based on a sparse approximation method. Besides imaging,
this approach can be used for denoising, and to determine the magnitude of
the scattering coefficients of the targets embedded in homogeneous media in
the presence of moderate noise levels. The method is formulated entirely in
the frequency domain, in the case where the Born approximation can be used
[10]. Relevant applications are radar imaging, exploration seismics,
nondestructive material testing, microwave breast imaging, ultrasound kidney
stones localization, and other acoustic inverse problems [1-9].

\section{Frequency response matrix}

We consider a system consisting of an array of $N$ transceivers (i.e. each
antenna is an emitter and a receiver) located at $x_{n}\in R^{D}$ $%
(n=1,...,N)$, and a collection of $M$ distinct scatterers (targets) with
scattering coefficients $\rho _{m}$, located at $y_{m}\in R^{D}$ $%
(m=1,...,M) $ (Fig. 1). Here, $D=1,2,3$ is the dimensionality of the space.
Also, we assume that the wave propagation is well approximated in the
space-frequency domain $(x,\omega )$ by the inhomogeneous Helmholtz
equation: 
\begin{equation}
\left[ \nabla ^{2}+k_{0}^{2}\eta ^{2}(x)\right] \psi (x,\omega )=-s(x,\omega
),
\end{equation}
where $\psi (x,\omega )$ is the wave amplitude produced by a localized
source $s(x,\omega )$, $k_{0}=2\pi \omega /c_{0}=2\pi /\lambda $ is the
wavenumber of the homogeneous background, with $\omega $ the frequency, $%
c_{0}$ the homogeneous background wave speed, and $\lambda $ the wavelength.
Here, $\eta (x)$ is the index of refraction: $\eta (x)=c_{0}/c(x)$, where $%
c(x)$ is the wave speed at location $x$. In the background we have $\eta
_{0}^{2}(x)=1$, while $\eta ^{2}(x)=1+\alpha (x)$, measures the change in
the wave speed at the scatterers location.

The fundamental solutions, or the Green functions, for this problem satisfy
the following equations: 
\begin{equation}
\left[ \nabla ^{2}+k_{0}^{2}\right] G_{0}(x,x^{\prime })=-\delta
(x-x^{\prime }),
\end{equation}
\begin{equation}
\left[ \nabla ^{2}+k_{0}^{2}\eta ^{2}(x)\right] G(x,x^{\prime })=-\delta
(x-x^{\prime }),
\end{equation}
for the homogeneous and inhomogeneous media, respectively. The expression of
the homogeneous Green function depends on the dimension of the space as
following: 
\begin{eqnarray}
G_{0}(x,x^{\prime }) &=&\frac{i}{2k_{0}}\exp (ik_{0}|x-x^{\prime }|),\quad
D=1, \\
G_{0}(x,x^{\prime }) &=&\frac{i}{4}H_{0}^{(1)}(k_{0}|x-x^{\prime }|),\quad
D=2, \\
G_{0}(x,x^{\prime }) &=&\frac{\exp (ik_{0}|x-x^{\prime }|)}{4\pi
|x-x^{\prime }|},\quad D=3,
\end{eqnarray}
where, $H_{0}^{(1)}(.)$ is the zero order Hankel function of the first kind.
Also, the fundamental solution $G(x,x^{\prime })$ for the inhomogeneous
medium can be written in terms of that for the homogeneous one $%
G_{0}(x,x^{\prime })$ as: 
\begin{equation}
G(x,x^{\prime })=G_{0}(x,x^{\prime })+k_{0}^{2}\int \alpha
(z)G_{0}(x,z)G(z,x^{\prime })dz.
\end{equation}
This is an implicit integral equation for $G(x,x^{\prime })$. Since the
scatterers are assumed to be pointlike, the regions with $\alpha (z)\neq 0$
are assumed to be finite, and included in compact domains $\Omega _{m}$
centered at $y_{m}$, $m=1,...,M$, which are small compared to the wavelength 
$\lambda $. Therefore we can write: 
\begin{equation}
\alpha (z,\omega )=\sum_{m=1}^{M}\rho _{m}(\omega )\delta (z-y_{m}),
\end{equation}
and consequently we obtain the Foldy-Lax equations: 
\begin{equation}
G(x,x^{\prime })\simeq G_{0}(x,x^{\prime })+\sum_{m=1}^{M}\rho _{m}(\omega
)G_{0}(x,y_{m})G(y_{m},x^{\prime }).
\end{equation}
If the scatterers are sufficiently far apart then we can neglect the
multiple scattering among the scatterers $(G(y_{m},x^{\prime })\simeq
G_{0}(y_{m},x^{\prime }))$ and we obtain the Born approximation of the
solution: 
\begin{equation}
G(x,x^{\prime })\simeq G_{0}(x,x^{\prime })+\sum_{m=1}^{M}\rho _{m}(\omega
)G_{0}(x,y_{m})G_{0}(y_{m},x^{\prime }).
\end{equation}
If $x$ corresponds to the receiver location $x_{i}$, and $x^{\prime }$
corresponds to the emitter location $x_{j}$, then we obtain: 
\begin{equation}
G(x_{i},x_{j})\simeq G_{0}(x_{i},x_{j})+H_{ij}(\omega ),
\end{equation}
where 
\begin{equation}
H_{ij}(\omega )=\sum_{m=1}^{M}G_{0}(x_{i},y_{m})\rho _{m}(\omega
)G_{0}(y_{m},x_{j}),\quad i,j=1,...,N,
\end{equation}
are the elements of the frequency response matrix $H(\omega )=[H_{ij}(\omega
)]$ [4-9]. The response matrix $H(\omega )$ is obviously a complex and
symmetric $N\times N$ matrix, since the same Green function is used in both
the transmission and the reception paths.

\section{Back-propagation and imaging}

An important step in computational time-reversal imaging is to determine the
frequency response matrix $H(\omega )$. This can be done by performing a
series of $N$ simple experiments, in which a single element of the array is
excited with a suitable signal $s$ and we measure the frequency response
between this element and all the other elements of the array [1-9]. In
general, given the Green function of the homogeneous media $%
G_{0}(x,x^{\prime })$, the general solution to the Helmholtz equation is the
convolution: 
\begin{equation}
\psi (x,\omega )=(G_{0}*s)(x)=\int G_{0}(x,x^{\prime })s(x^{\prime },\omega
)dx^{\prime }.
\end{equation}
Thus, if the $j$ antenna emits a signal $s_{j}$ then, using the convolution
theorem in the Fourier domain, the field produced at the location $r$ is $%
G_{0}(r,x_{j})s_{j}$. If this field is incident on the $m$-th scatterer, it
produces at $r$ the scattered field $G_{0}(r,y_{m})\rho _{m}(\omega
)G_{0}(r,x_{j})s_{j}$. Thus, the total wave field at location $r$, due to a
pulse emitted by a single element at $x_{j}$ and scattered by the $M$
targets can be expressed as: 
\begin{equation}
\psi (r,\omega )=\sum_{m=1}^{M}G_{0}(r,y_{m})\rho _{m}(\omega
)G_{0}(y_{m},x_{j})s_{j}.
\end{equation}
If this field is measured at the $i$-th antenna we obtain: 
\begin{equation}
\psi (x_{i},\omega )=\sum_{m=1}^{M}G_{0}(x_{i},y_{m})\rho _{m}(\omega
)G_{0}(y_{m},x_{j})s_{j}=H_{ij}(\omega )s_{j}.
\end{equation}
Thus, the response matrix can be rewritten as: 
\begin{equation}
H(\omega )=\Gamma (\omega )R(\omega )\Gamma ^{T}(\omega ),
\end{equation}
where 
\begin{equation}
R(\omega )=diag\{\rho _{1}(\omega ),...,\rho _{M}(\omega )\},
\end{equation}
and 
\begin{equation}
\Gamma (\omega )=\left[ 
\begin{array}{llll}
g(y_{1},\omega ) & g(y_{1},\omega ) & ... & g(y_{M},\omega )
\end{array}
\right] .
\end{equation}
Here, $g(y_{m},\omega )$ is the Green function vector associated with the $m$%
-th scatterer 
\begin{equation}
g(y_{m},\omega )=\left[ 
\begin{array}{llll}
G_{0}(x_{1},y_{m},\omega ) & G_{0}(x_{2},y_{m},\omega ) & ... & 
G_{0}(x_{N},y_{m},\omega )
\end{array}
\right] ^{T}.
\end{equation}
In general, the Green function vector defined as: 
\begin{equation}
g(r,\omega )=\left[ 
\begin{array}{llll}
G_{0}(x_{1},r,\omega ) & G_{0}(x_{2},r,\omega ) & ... & G_{0}(x_{N},r,\omega
)
\end{array}
\right] ^{T},
\end{equation}
expresses the response at each array element due to a single pulse emitted
from $r$.

In the formulation of time-reversal imaging one forms the self-adjoint
matrix [6]: 
\begin{equation}
K(\omega )=H^{*}(\omega )H(\omega )=\overline{H}(\omega )H(\omega ),
\end{equation}
where the star denotes the adjoint and the bar denotes the complex conjugate
($H^{*}=\overline{H}$, since $H$ is symmetric). $\overline{H}$ is the
frequency-domain version of a time-reversed response matrix, thus $K(\omega
) $ corresponds to performing a scattering experiment, time-reversing the
received signals and using them as input for a second scattering experiment.
Therefore, time-reversal imaging relies on the assumption that the Green
function can be always calculated.

As long as the number of antenna elements exceeds the number of scatterers, $%
M<N$, the matrix $K(\omega )$ is rank deficient and it has only $M$ non-zero
eigenvalues, with the corresponding eigenvectors $V_{m}(\omega )$, $%
m=1,...,M $. When the scatterers are well resolved, the columns of the
matrix $\Gamma (\omega )$ are approximately orthogonal to each other, and
the eigenvectors can be back-propagated as $g^{T}(r,\omega )V_{m}(\omega )$,
and consequently the radiated wavefields focus at target locations. Thus,
each eigenvector can be used to locate a single scatterer. For example, let
us consider a scenario consisting of $N=100$ transceivers, separated by 
$d=\lambda /2$, and located at $x_{n}=\left[ 
\begin{array}{ll}
0 & n\lambda /2+a/2-N\lambda /4
\end{array}
\right] ^{T}$. Also, there are two targets, $M=2$, with the scattering
coefficients $\rho _{1,2}=1$, situated at $y_{1}=\left[ 
\begin{array}{ll}
0.65 & 0.25
\end{array}
\right] ^{T}a$, and respectively $y_{2}=\left[ 
\begin{array}{ll}
0.80 & 0.75
\end{array}
\right] ^{T}a$. Here, $a=100\lambda $ is the side of the imaging area, and
the computational image grid is set to $L\times L=200\times 200$ pixels. In
Figure 2 we give the first two back-propagated eigenvectors and the computed
time-reversal image. One can see that since the targets are well separated
the computed time-reversal image is almost a perfect superposition of the
two independently back-propagated eigenvectors.

\section{Subspace-based imaging}

The above result does not apply to the case of poorly-resolved targets. In
this case, the eigenvectors of $K(\omega )$ are linear combinations of the
target Green function vectors $g(y_{m},\omega )$. Thus, back-propagating one
of these eigenvectors generates a linear combination of wavefields, each
focused on a different target location. The subspace-based algorithms, based
on the multiple signal classification (MUSIC) method, can be used in this
more general situation [7-9]. The signal subspace method assumes that the
number $M$ of point targets in the medium is lower than the number of
transceivers $N$, and the general idea is to localize multiple sources by
exploiting the eigenstructure and the rank deficiency of the response matrix 
$H(\omega )$.

The singular value decomposition of the symmetrical matrix $H(\omega )$ is
given by: 
\begin{equation}
H(\omega )=U(\omega )\Lambda (\omega )V^{*}(\omega ),
\end{equation}
where $U(\omega )$ and $V(\omega )$ are the $N\times N$ orthogonal matrices
corresponding to the left and right singular vectors ($V(\omega )=\overline{U%
}(\omega )$, since $H(\omega )$ is symmetric). The singular value matrix $%
\Lambda (\omega )$ is diagonal $\Lambda (\omega )=diag\{\lambda _{1}(\omega
),...,\lambda _{N}(\omega )\}$, and since $H(\omega )$ is rank-deficient,
all but the first $M$ singular values vanish: 
\begin{eqnarray}
\quad \lambda _{i}(\omega ) &\neq &0,\quad i=1,...,M, \\
\lambda _{j}(\omega ) &=&0,\quad j=M+1,...,N.
\end{eqnarray}
Therefore, the first $M$ columns (singular vectors) of $V(\omega )$ span the
same subspace $\sigma $ as the columns of $\Gamma (\omega )$, while the
remaining $N-M$ columns span the null-subspace $\nu $ of $\Gamma (\omega )$.
Thus, by partitioning $V(\omega )$ as: 
\begin{equation}
V(\omega )=\left[ 
\begin{array}{ll}
V_{\sigma }(\omega ) & V_{\nu }(\omega )
\end{array}
\right]
\end{equation}
where $V_{\sigma }(\omega )$ has the first $M$ columns and $V_{\nu }(\omega
) $ has the remaining $N-M$ columns, one can write the signal space as a
direct sum $\sigma \oplus \nu $, where the essential signal-subspace $\sigma 
$ is orthogonal to the null-subspace $\nu $. It follows immediately that: 
\begin{equation}
V_{\nu }^{*}(\omega )H(\omega )=0,
\end{equation}
and therefore 
\begin{equation}
V_{\nu }^{*}(\omega )g(r,\omega )=0,
\end{equation}
for any $\omega $. Therefore, the target locations must correspond to the peaks
in the MUSIC pseudo-spectrum for any $\omega $: 
\begin{equation}
P_{MUSIC}(r,\omega )=\left\| V_{\nu }^{*}(\omega )g(r,\omega )\right\| ^{-2},
\end{equation}
where $g(r,\omega )$ is the free-space Green function vector. Thus, one can
form an image of the scatterers by plotting, at each point $r$, the quantity 
$P_{MUSIC}(r,\omega )$. The resulting plot will have large peaks at the
locations of the scatterers. For example, let us consider a two dimensional
scenario, consisting of $N=100$ linearly distributed transceivers, separated 
by $d=\lambda /2$ and located at $x_{n}=\left[ 
\begin{array}{ll}
0 & n\lambda /2+a/2-N\lambda /4
\end{array}
\right] ^{T}$, where $a=100\lambda $ is the side of the imaging area. Also,
there are $M=5$ targets with the scattering coefficients $\rho _{m}=10$, $%
m=1,...,M$. The computational image grid is set to $L\times L=200\times 200$
pixels. We consider two cases, one without noise, and the second one with
Gaussian noise added to the elements of the response matrix $H_{ij}(\omega )$%
. The noise level was set such that the signal to noise ratio (SNR) is $%
SNR=2 $. SNR compares the level of a desired signal to the level of
background noise. The higher the ratio, the less obtrusive the background
noise is. SNR measures the power ratio between a signal and the background
noise: 
\begin{equation}
SNR=P_{signal}/P_{noise}=(A_{signal}/A_{noise})^{2},
\end{equation}
where $P$ is average power and $A$ is root mean square (RMS) amplitude. In
Figure 3 we give the results obtained with the MUSIC algorithm in both
cases. One can see that the MUSIC pseudo-spectrum provides a better
resolution and separation of the targets, comparing to the back-propagation
method.

\section{Sparse approximation problem}

The MUSIC algorithm provide very good resolution and separation of targets,
however it cannot be used to quantify the properties of the targets, such as
the magnitude of their scattering coefficients. Here we show that, with a
little more computational effort, we can also determine the scattering
coefficients using a sparse image reconstruction approach.

Let us assume that the imaging domain is discretized as a grid of $L^{D}$
voxels (pixels), and $r_{l}$, $l=1,...,L^{D}$, gives the position of each
voxel. Also, we assume that $\widetilde{\rho }_{l}(\omega )$, $l=1,...,L^{D}$%
, is the scattering factor associated with each voxel in the imaging domain.
The goal is to find a matrix 
\begin{equation}
\widetilde{H}(\omega )=\sum_{l=1}^{L^{D}}\widetilde{\rho }_{l}(\omega
)g(r_{l},\omega )g(r_{l},\omega )^{T},
\end{equation}
which best approximates the response matrix $H(\omega )$: 
\begin{equation}
\widetilde{H}(\omega )\simeq H(\omega ).
\end{equation}
The above equation can be rewritten as: 
\begin{equation}
\Phi (\omega )\widetilde{\rho }(\omega )\simeq \Theta (\omega ),
\end{equation}
where $\widetilde{\rho }(\omega )$ is the unknown $L^{D}$-dimensional
vector: 
\begin{equation}
\widetilde{\rho }(\omega )=[\widetilde{\rho }_{1}(\omega ),...,\widetilde{%
\rho }_{L^{D}}(\omega )]^{T}.
\end{equation}
The $N^{2}$-dimensional vector $\Theta (\omega )$ is obtained by stacking
the columns 
\[
H_{n}(\omega )=\left[ 
\begin{array}{lll}
H_{1n}(\omega ) & ... & H_{Nn}(\omega )
\end{array}
\right] ^{T} 
\]
of the response matrix $H(\omega )$: 
\begin{equation}
\Theta (\omega )=\left[ 
\begin{array}{lllllll}
H_{11}(\omega ) & ... & H_{N1}(\omega ) & ... & H_{1N}(\omega ) & ... & 
H_{NN}(\omega )
\end{array}
\right] ^{T}.
\end{equation}
Also, $\Phi (\omega )$ is a matrix with $N^{2}$ rows and $L^{D}$ columns.
Each column $\Phi _{l}(\omega )$, $l=1,...,L^{D}$, is obtained in a similar
way, by stacking the columns of the $N\times N$ matrix $g(r_{l},\omega
)g(r_{l},\omega )^{T}$.

The above system of equations is underdetermined, since the number of
scattering targets $M$ is much smaller than $L^{D}$. The common approach to
find a solution is to consider the equivalent $l_{2}$-optimization problem
[11]: 
\begin{equation}
\widetilde{\rho }(\omega )=\arg \min_{\widetilde{\rho }(\omega )}\left\| 
\widetilde{\rho }(\omega )\right\| _{2}\quad s.t.\quad \Phi (\omega )%
\widetilde{\rho }(\omega )=\Theta (\omega ),
\end{equation}
where $\left\| \widetilde{\rho }(\omega )\right\| _{2}=\sqrt{%
\sum_{l=1}^{L^{D}}\left| \widetilde{\rho }_{l}(\omega )\right| ^{2}}$, is
the Euclidean norm. In this case, the unique solution which minimizes the $%
l_{2}$-norm, is given by: 
\begin{equation}
\widetilde{\rho }(\omega )=\Phi ^{\dagger }(\omega )\Theta (\omega ),
\end{equation}
where 
\begin{equation}
\Phi ^{\dagger }(\omega )=[\Phi ^{*}(\omega )\Phi (\omega )]^{-1}\Phi
^{*}(\omega ),
\end{equation}
is the More-Penrose pseudoinverse of the matrix $\Phi (\omega )$. However,
in this case all the coefficients of the solution $\widetilde{\rho }(\omega
) $ are non-zero, and therefore, this solution is not in agreement with the
fact that the imaging region is sparse, i.e. the number of scattering
targets is $M\ll L^{D}$. Therefore, this is not the correct solution of the
approximation problem. In fact, we would like to find the minimum number of
columns $\Phi _{l}(\omega )$ of $\Phi (\omega )$ which approximate the data
vector $\Theta (\omega )$. This is a sparse approximation problem, and as a
measure of sparsity we consider the $l_{0}$ norm of $\widetilde{\rho }%
(\omega )$, $\left\| \widetilde{\rho }(\omega )\right\| _{0}$, which simply
counts the number of nonzero coefficients in the vector $\widetilde{\rho }%
(\omega )$. These non-zero coefficients will give the position, and their
magnitude will reflect the value, of the scattering coefficients of the
targets in the $L^{D}$ imaging region. Thus, the sparsest representation
requires the solution of the $l_{0}-$optimization problem [12]: 
\begin{equation}
\widetilde{\rho }(\omega )=\arg \min_{\widetilde{\rho }(\omega )}\left\| 
\widetilde{\rho }(\omega )\right\| _{0}\quad s.t.\quad \Phi (\omega )%
\widetilde{\rho }(\omega )=\Theta (\omega ).
\end{equation}
Unfortunately, this combinatorial optimization problem is NP-hard to solve,
requiring the enumeration of all possible collections of columns in $\Phi
(\omega )$ and searching for the smallest collection which best approximates
the data vector $\Theta (\omega )$. An alternative is the convexification of
the objective function, which is obtained by replacing the $l_{0}$ norm with
the $l_{1}$ norm: $\left\| \widetilde{\rho }(\omega )\right\|
_{1}=\sum_{l=1}^{L^{D}}\left| \widetilde{\rho }_{l}(\omega )\right| $. The
resulting $l_{1}$-optimization problem: 
\begin{equation}
\widetilde{\rho }(\omega )=\arg \min_{\widetilde{\rho }(\omega )}\left\| 
\widetilde{\rho }(\omega )\right\| _{1}\quad s.t.\quad \Phi (\omega )%
\widetilde{\rho }(\omega )=\Theta (\omega )
\end{equation}
is known as Basis Pursuit (BP), and it can be solved using linear
programming techniques whose computational complexities are polynomial [12].
The BP method recasts the $l_{1}$-problem as a linear program, and it has
been shown that because of the nondifferentiability of the $l_{1}$ norm,
this optimization problem leads to unique sparse solutions. However, the BP
approach requires the solution of a very large convex, nonquadratic
optimization problem, and therefore still suffers from high computational
complexity. For example in a three dimensional problem, $D=3$, with $N=100$
transceivers and a discretization grid with $L=100$, the resulted
dimensionality of the $\Phi (\omega )$ matrix is: $10^{4}\times 10^{6}$, and
the number of unknowns in the vector $\widetilde{\rho }(\omega )$ is $10^{6}$%
. As an alternative, here we consider a heuristic approach based on
iterative greedy algorithms, which also have been proven to give good
approximative solutions to the sparse reconstruction problem.

\section{Greedy algorithm for sparse approximation}

Matching Pursuit (MP) is a general procedure to compute adaptive signal
representations and to extract the signal structure in a given
time-frequency dictionary [13]. Also, it has been shown that the MP
algorithm can be used to obtain (approximative) sparse solutions of the $%
l_{0}$-optimization problem [14-16]. Although the MP algorithm is non-linear,
it maintains an energy conservation which guarantees its convergence.

In the case of computational time-reversal imaging, the elements of the
time-frequency dictionary are given by the columns $\Phi _{l}(\omega )$ of
the matrix $\Phi (\omega )$. Using this dictionary, the data vector $\Theta
(\omega )$ can be represented as: 
\begin{equation}
\Theta (\omega )=\sum_{l=1}^{L^{D}}\widetilde{\rho }_{l}(\omega )\Phi
_{l}(\omega ).
\end{equation}
The vector $\Theta (\omega )$ can be decomposed into: 
\begin{equation}
\Theta (\omega )=\left[ \Phi _{l}^{*}(\omega )\Psi (\omega )\right] \left\|
\Phi _{l}(\omega )\right\| ^{-2}\Phi _{l}(\omega )+\Psi (\omega ),
\end{equation}
where $\Psi (\omega )$ is the residual vector after approximating $\Theta
(\omega )$ in direction of $\Phi _{l}(\omega )$. Since $\Phi _{l}(\omega )$
and $\Psi (\omega )$ are orthogonal, we have: 
\begin{equation}
\left\| \Psi (\omega )\right\| ^{2}=\left\| \Theta (\omega )\right\|
^{2}-\left| \Phi _{l}^{*}(\omega )\Theta (\omega )\right| ^{2}\left\| \Phi
_{l}(\omega )\right\| ^{-2},
\end{equation}
and in order to minimize $\left\| \Psi (\omega )\right\| $ we must choose
the column $\Phi _{l}(\omega )$, such that $\left| \Phi _{l}^{*}(\omega
)\Theta (\omega )\right| \left\| \Phi _{l}(\omega )\right\| ^{-1}$ is
maximum. Thus, starting from an initial approximation $\widetilde{\rho }%
(\omega )=0$ and a residual $\Psi (\omega )=\Theta (\omega )$, the algorithm
uses an iterative greedy strategy to pick the columns $\Phi _{l}(\omega )$
of $\Phi (\omega )$ that are the most strongly correlated with the residual.
Then, successively their contribution is subtracted from the residual, which
this way can be made arbitrarily small. The pseudo-code of the MP algorithm
is:

\begin{enumerate}
\item  Initialize the variables: 
\begin{equation}
T,t\leftarrow 1,\widetilde{\rho }(\omega )\leftarrow 0,\Psi (\omega
)\leftarrow \Theta (\omega ).
\end{equation}

\item  Find $l$ such that: 
\begin{equation}
l=\arg \max_{l=1,...,L^{D}}\left| \Phi _{l}^{*}(\omega )\Psi (\omega
)\right| \left\| \Phi _{l}(\omega )\right\| ^{-1}.
\end{equation}

\item  Update the estimate of the corresponding coefficient, the residual,
and the iteration counter: 
\begin{equation}
\widetilde{\rho }_{l}(\omega )\leftarrow \widetilde{\rho }_{l}(\omega
)+\left[ \Phi _{l}^{*}(\omega )\Psi (\omega )\right] \left\| \Phi
_{l}(\omega )\right\| ^{-2},
\end{equation}
\begin{equation}
\Psi (\omega )\leftarrow \Psi (\omega )-\left[ \Phi _{l}^{*}(\omega )\Psi
(\omega )\right] \left\| \Phi _{l}(\omega )\right\| ^{-2}\Phi _{l}(\omega ),
\end{equation}
\begin{equation}
t\leftarrow t+1.
\end{equation}

\item  If $\left\| \Psi (\omega )\right\| ^{2}<\varepsilon \left\| \Theta
(\omega )\right\| ^{2}$ or $t>T$ then terminate and return $\widetilde{\rho }%
(\omega )$. Otherwise go to 2.
\end{enumerate}

The stopping criterion in the step 4 requires the residual to be smaller
than some fraction $0<\varepsilon \ll 1$ of the data vector. Also, the
computation stops if the number of iterations exceed the maximum number
allowed $T$. Although the asymptotic convergence of MP algorithm can be
easily proven, the resulting approximation after any finite number of steps
will in general be suboptimal. In the case of noise, the MP algorithm is
used to obtain an approximative sparse solution by simply stopping the
iteration when the projection of the residual on the chosen direction $\Phi
_{l}(\omega )$ becomes smaller than a threshold $0<\tau <1$ [14-16]. Thus, in
the case of noise, the MP algorithm is modified simply by replacing the
stopping condition with: $\left| \Phi _{l}^{*}(\omega )\Psi (\omega )\right|
^{2}\left\| \Phi _{l}(\omega )\right\| ^{-2}\left\| \Psi (\omega )\right\|
^{-2}<\tau $. After the computation is finished, the image is formed by
plotting $\widetilde{\rho }_{l}(\omega )$ at location $r_{l}$, $l=1,...,L^{D}
$. We should note that the algorithm works in both cases, when $\rho
_{l}(\omega )$ is complex, or when only its magnitude $\left| \rho
_{l}(\omega )\right| $ is given. In the later case, one should plot the
absolute values of the computed coefficients $\left| \widetilde{\rho }_{l}(\omega
)\right| $.

\section{Implementation and numerical results}

It is important to note that one can implement the MP algorithm such that
the elements of the matrix $\Phi $ do not need to be stored. In fact, one
can compute the columns $\Phi _{l}$ of $\Phi $ at every step of the
algorithm, since the Green function is known. Thus, the algorithm requires
only operations with vectors of length $N^{2}$, which is feasible on
standard personal computers. However, since the number of vector-vector
multiplications per iteration step is high, $L^{2}$, a parallel
implementation of the algorithm is desirable. We have implemented the MP
algorithm for the NVIDIA GPU (Graphics Processing Unit) platform. Recently,
NVIDIA has released a general purpose oriented API for its graphics
hardware, called CUDA [17]. In addition, NVIDIA has developed CUBLAS which
is a GPU optimized version of BLAS library (Basic Linear Algebra
Subroutines) built on top of CUDA [18]. The newly developed GPUs now include
fully programmable processing units that follow a stream programming model
and support vectorized single and double precision floating-point
operations. For example, the CUDA computing environment provides a standard
C like language interface to the NVIDIA GPUs. The computation is distributed
into sequential grids, which are organized as a set of thread blocks. The
thread blocks are batches of threads that execute together, sharing local
memories and synchronizing at specified barriers. CUBLAS library provides
functions for: (i) creating and destroying matrix and vector objects in GPU
memory; (ii) transferring data from CPU mainmemory to GPU memory; (iii)
executing BLAS on the GPU; (iv) transferring data from GPU memory back to
the CPU mainmemory. CUBLAS defines a set of fundamental operations on
vectors and matrices which can be used to create optimized higher-level
linear algebra functionality: (i) Level 1 BLAS perform scalar, vector and
vector-vector operations; (ii) Level 2 BLAS perform matrix-vector
operations; (iii) Level 3 BLAS perform matrix-matrix operations. However, in
its current version CUBLAS does not offer direct support for operations
involving vectors and matrices of complex numbers. In the case of the MP
algorithm, one can easily overcome this drawback by storing separately the
real and imaginary part of the vectors and matrices. This way, complex
vector-vector and matrix-vector operations can be reduced to operations
involving only real numbers. The parallel implementation of the MP algorithm
requires only Level 1 or Level 2 (if $\Phi $ is stored) BLAS operations. Our
numerical tests have shown that the parallel GPU implementation versus a
standard CPU BLAS implementation of the MP algorithm reaches a speed up of a
maximum of 31 times in single precision and respectively 21 times in double
precision [19]. 

Let us now consider a two dimensional scenario, consisting of $N=100$
linearly distributed transceivers, separated by $d=\lambda /2$ and located 
at $x_{n}=\left[ 
\begin{array}{ll}
0 & n\lambda /2+a/2-N\lambda /4
\end{array}
\right] ^{T}$, where $a=100\lambda $ is the side of the imaging area. The
computational image grid was also set to $L\times L=100\times 100$. This
scenario will generate a dictionary $\Phi $ of size $N^{2}\times
L^{2}=10^{4}\times 10^{4}$ and an unknown vector $\widetilde{\rho }$ of size 
$N^{2}=10^{4}$. The number of targets is set to $M=5$ and their position is
randomly generated in the imaging area. The scattering coefficients are
generated randomly from a uniform distribution such that $1\leq \rho
_{m}\leq 10$, $m=1,...,M$. We consider both cases, with and without noise.
The signal to noise level is set to $SNR=2$, as for the described MUSIC
algorithm case.

First, let us discuss the case without noise. The threshold parameter and
the maximum number of iterations were set to $\varepsilon =10^{-4}$, and
respectively $T=N$. In Figure 4 we give the initial and the computed
arrangement of the targets, and their initial and computed values of the
scattering coefficients. One can see that the agreement between the initial
values and the computed ones is very good, and the MP algorithm can solve
the problem almost perfectly. In Figure 5 we give the real and imaginary
part of the initial and reconstructed response matrix. The error is less
than $1\%$ in both cases. Also, Figure 6 shows the computed image using the
MP algorithm. One can see that the peaks are very sharp and their position
and amplitude reflects correctly the initial position and the magnitude of
the scattering coefficients. This example shows that the MP algorithm can
solve almost perfectly (in the limits of the image resolution) the
computational time-reversal imaging problem if the response matrix is not
affected by noise.

Let us consider the case with noise, using the same arrangement of the
targets, with the same values for the scattering coefficients. In this case
the threshold $\tau $ plays a very important role in the MP algorithm. We
consider two different values $\tau =10^{-3}$ and $\tau =10^{-2}$. In Figure
7 we give the real and the imaginary part of the initial and reconstructed
response matrix. One can see that the reconstructed response matrix contains
a much lower amount of noise than the original response matrix. Also, by
increasing the value of $\tau $, the amount of noise in the reconstructed
response matrix decreases dramatically. This can be seen also on the
computed images, given in Figure 8. The amount of noise in the computed
images is very small, given the $SNR=2$ level in the perturbed response
matrix. Also, the position of the peaks, corresponding to the targets, and
their magnitude are still very well maintained. This means that the sparse
approximation, given by the MP algorithm, and the threshold parameter $\tau $ 
, acts as a denoising method, and respectively as a denoising parameter, 
and it can be succesfully used in computational time-reversal imaging. 

\section{Conclusion}

We have presented a sparse approximation method for computational
time-reversal imaging. The method is formulated entirely in the frequency
domain, and it is based on an adapted version of the matching pursuit
algorithm, which can be successfully used to compute an accurate sparse
approximation of the frequency response matrix. This approach can be used
for denoising the computed time-reversal images, and to determine the
magnitude of the scattering coefficients of the targets embedded in
homogeneous media, in the presence of moderate to high noise levels. Also,
in comparison to the back-propagation and the null subspace projection
methods, the described approach provides a better resolution. However, the
sparse approximation method is computationally more expensive than the
traditional approaches.

\pagebreak

\clearpage
\begin{figure}
\centering
\includegraphics[width=7cm]{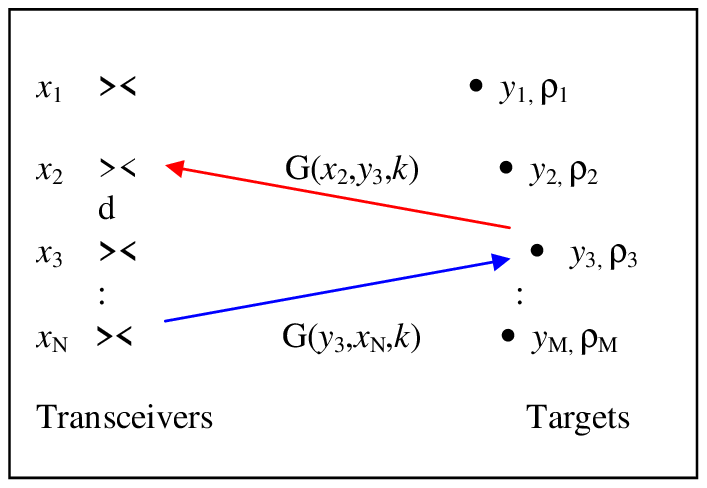}
\caption{\label{Fig1} Geometry for the time-reversal imaging experiment, containing $N$
transceivers and $M$ scattering targets.}
\end{figure}

\clearpage
\begin{figure}
\centering
\includegraphics[width=11cm]{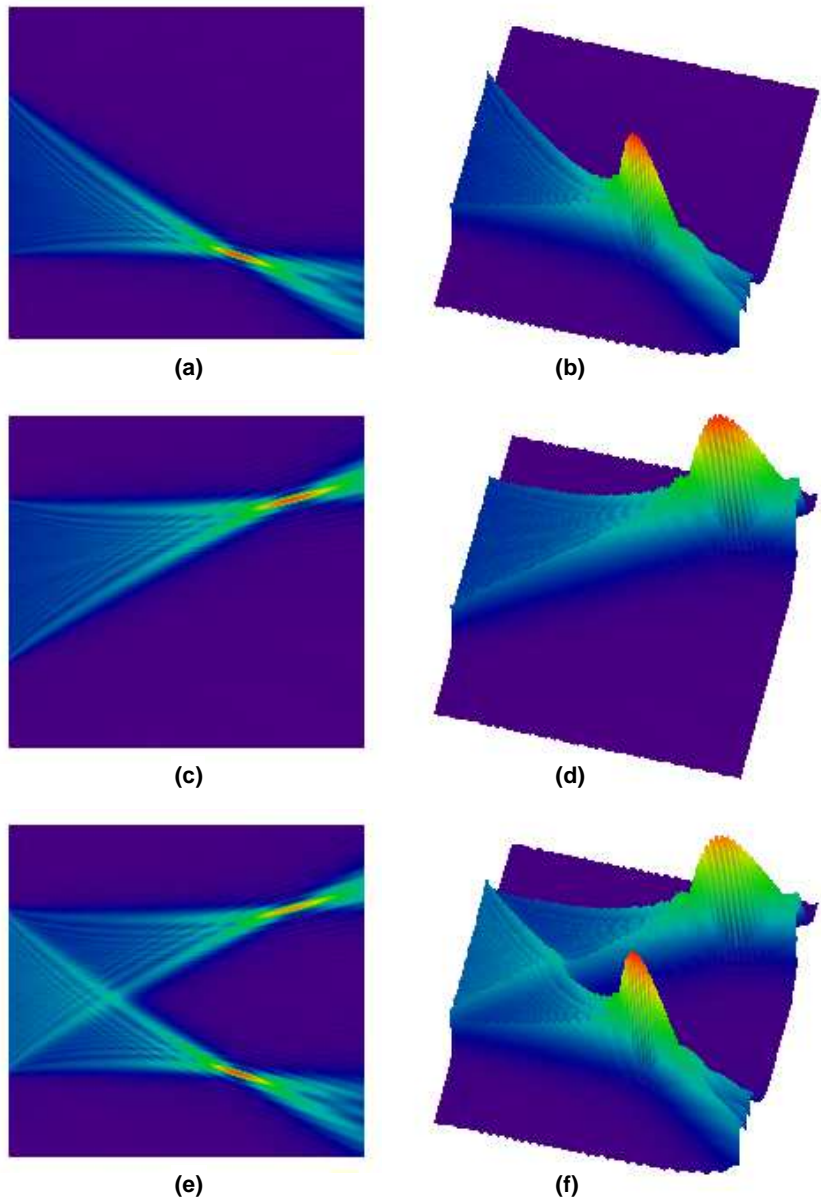}
\caption{\label{Fig2} Time-reversal wavefields in the case of two targets: (a-b) the first
eigenvector; (c-d) the second eigenvector; (e-f) the computed time-reversal
image.}
\end{figure}

\clearpage
\begin{figure}
\centering
\includegraphics[width=11cm]{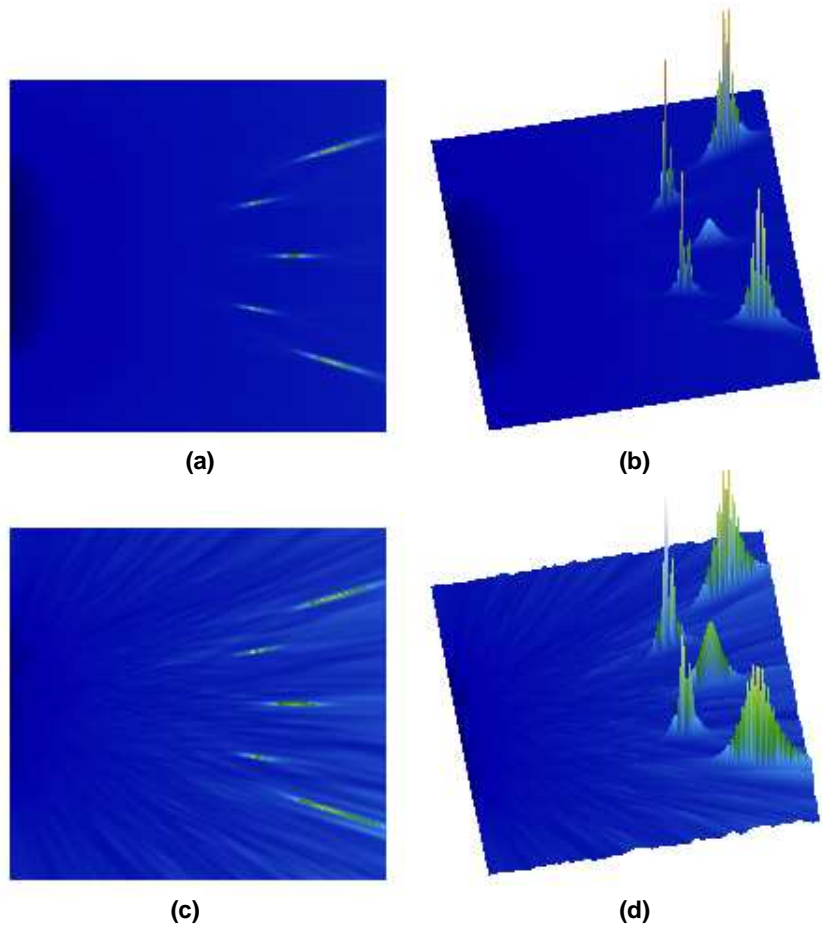}
\caption{\label{Fig3} The MUSIC pseudo-spectrum in case of $M=5$ targets: (a-b) without
noise; (c-d) with noise, $SNR=2$.}
\end{figure}

\clearpage
\begin{figure}
\centering
\includegraphics[width=11cm]{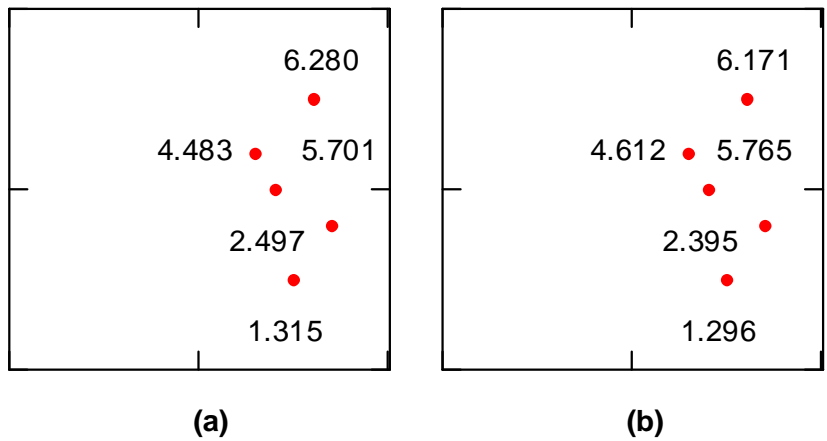}
\caption{\label{Fig4} Sparse approximation of time-reversal image in case of $M=5$
targets: (a) the initial location of the targets and their scattering
coefficients; (b) the computed location of the targets and their scattering
coefficients.}
\end{figure}

\clearpage
\begin{figure}
\centering
\includegraphics[width=11cm]{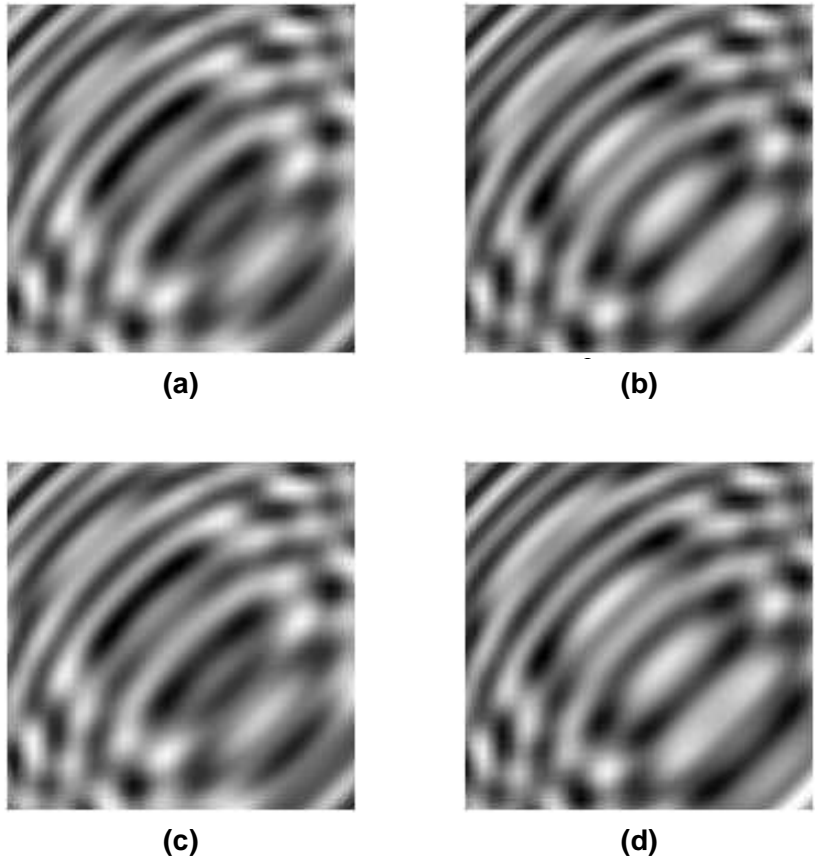}
\caption{\label{Fig5} Sparse approximation of time-reversal image in case of $M=5$
targets, without noise: (a-b) the real and imaginary part of the
initial response matrix; (c-d) the real and imaginary part of the
computed response matrix.}
\end{figure}

\clearpage
\begin{figure}
\centering
\includegraphics[width=11cm]{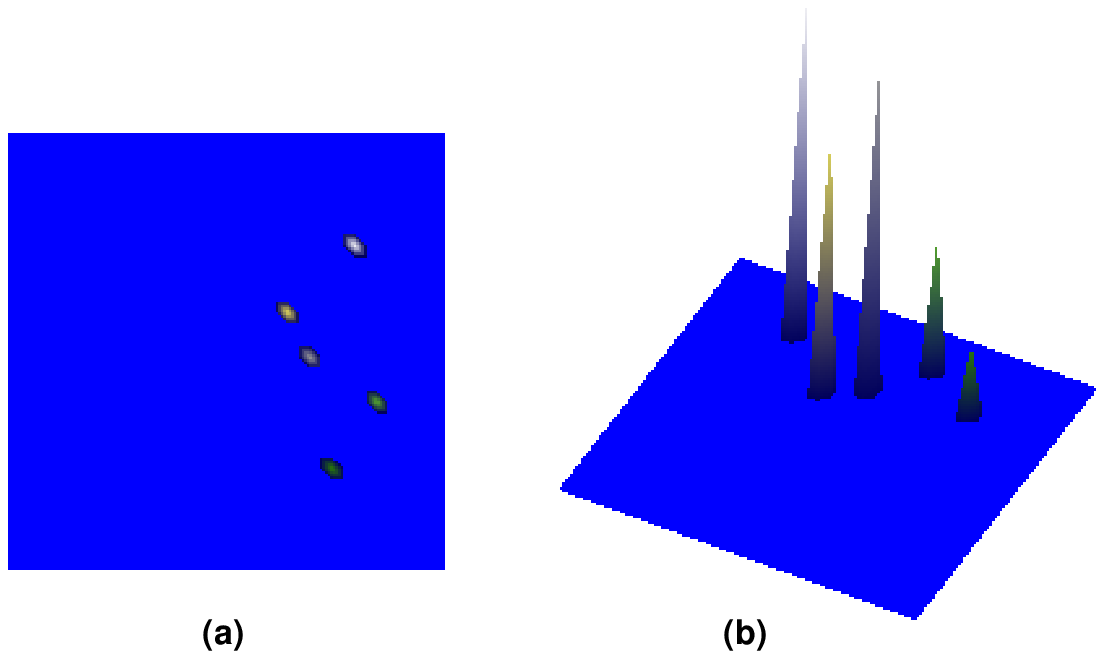}
\caption{\label{Fig6} The computed time-reversal image, using the sparse approximation
approach, in case of $M=5$ targets, without noise.}
\end{figure}

\clearpage
\begin{figure}
\centering
\includegraphics[width=11cm]{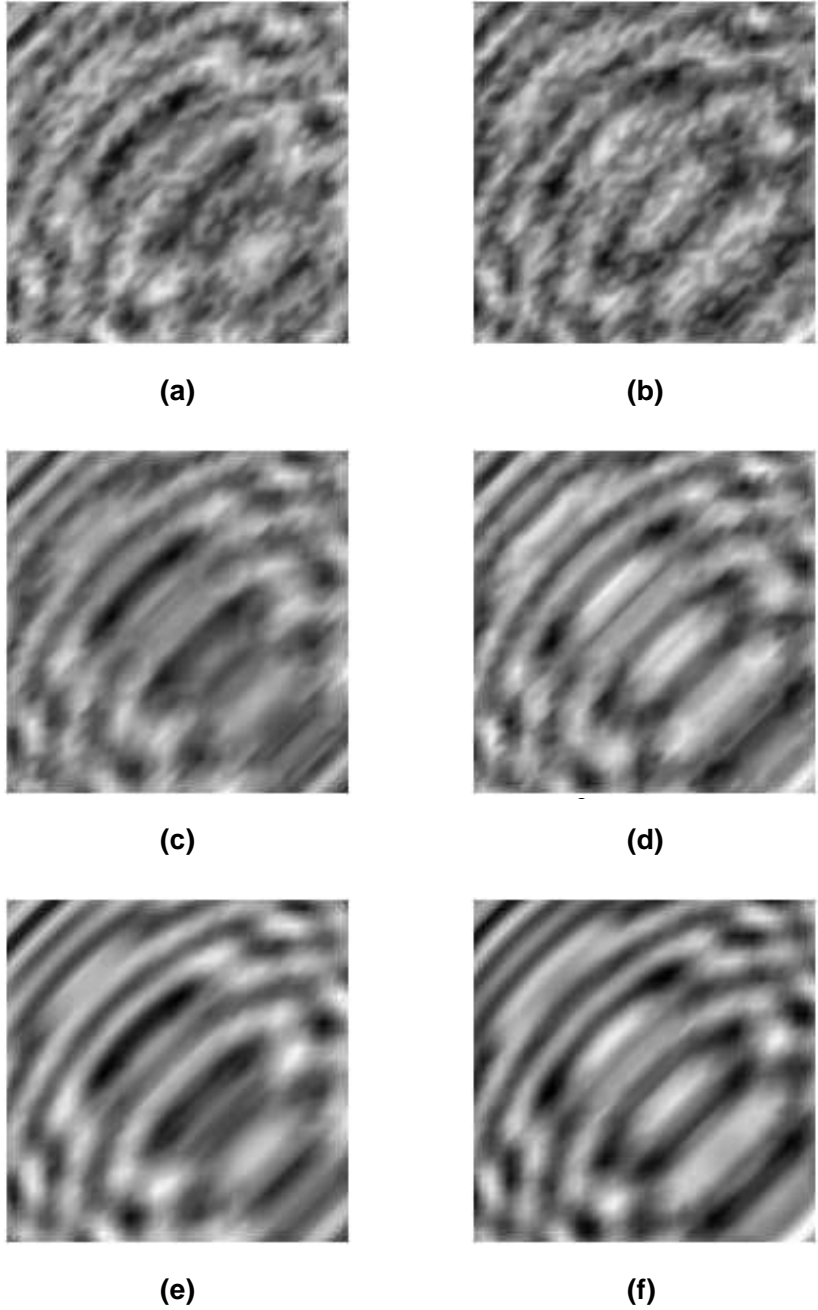}
\caption{\label{Fig7} Sparse approximation of time-reversal image in case of $M=5$
targets, with noise: (a-b) the real and imaginary part of the
initial response matrix with the noise level $SNR=2$; (c-d) the real and
imaginary part of the computed response matrix with $\tau =10^{-3}$;
(e-f) the real and imaginary part of the computed response matrix
with $\tau =10^{-2}$.}
\end{figure}

\clearpage
\begin{figure}
\centering
\includegraphics[width=11cm]{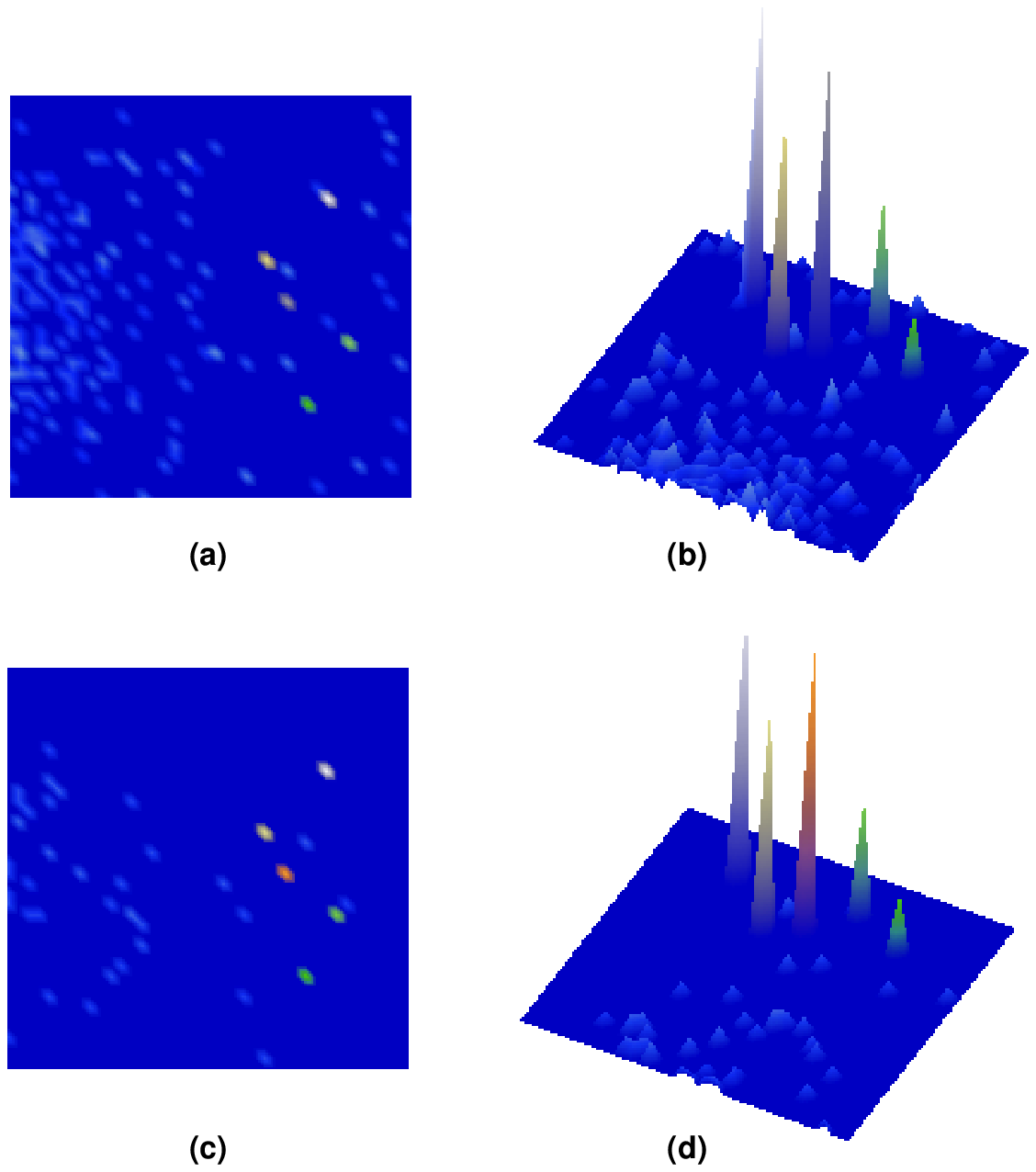}
\caption{\label{Fig8} The computed time-reversal image, using the sparse approximation
approach, in case of $M=5$ targets, with noise ($SNR=2$): (a-b) $\tau
=10^{-3}$; (c-d) $\tau =10^{-2}$.}
\end{figure}


\begin{thebibliography}{99}
\bibitem{1}  L. Borcea, G. Papanicolaou, C. Tsogka, J. Berryman, Imaging and
time reversal in random media, Inverse Problems, 18 (2002) 1247.

\bibitem{2}  M. Fink, D. Cassereau, A. Derode, C. Prada, P. Roux, M. Tanter,
J.-L. Thomas and F. Wu, Reports on Progress in Physics 63 (2000) 1933.

\bibitem{3}  C. Prada, E. Kerbrat, D. Cassereau, M. Fink, Time reversal
techniques in ultrasonic nondestructive testing of scattering media, Inverse
Problems, 18 (2002) 1761.

\bibitem{4}  C. Prada, L. Thomas, M. Fink, The Iterative Time Reversal
Process: Analysis of the Convergence, Journal of the Acoustical Sociefy of
America, 97 (1995) 62.

\bibitem{5}  C. Prada, M. Fink, Eigenmodes of the time reversal operator: A
solution to selective focusing in multiple-target media, Wave Motion, 20
(1994) 151.

\bibitem{6}  C. Prada, S. Manneville. D. Spoliansky, M. Fink, Decomposition
of the Time Reversal Operator: Detection and Selective Focusing on Two
Scatterers, Journal of the Acoustical Society of America, 99 (1996) 2067.

\bibitem{7}  F.K. Gruber, E.A. Marengo, A.J. Devaney, Timereversal imaging
with multiple signal classification considering multiple scattering between
the targets, Journal of the Acoustical Society of America, 115 (2004) 3042.

\bibitem{8}  E.A. Marengo, F.K. Gruber, Subspace-Based Localization and
Inverse Scattering of Multiply Scattering Point Targets, EURASIP Journal on
Advances in Signal Processing, (2007) Article ID 17342.

\bibitem{9}  H. Lev-Ari, A. J. Devaney, The time-reversal technique
reinterpreted: Subspace-based signal processing for multi-static target
location, IEEE Sensor Array and Multichannel Signal Processing Workshop,
Cambridge (MA), USA, (2000) 509.

\bibitem{10}  J. H. Taylor, Scattering Theory, Wiley, New York, 1972.

\bibitem{11}  G. H.Golub, C. F. Van Loan, Matrix Computations. Johns Hopkins
University Press, Baltimore, 1996.

\bibitem{12}  S. S. Chen, D. L. Donoho, M. A. Saunders, Atomic decomposition
by basis pursuit, SIAM Journal of Scientific Computing 20 (1998) 3361.

\bibitem{13}  S. Mallat, Z. Zhang, Matching pursuit in a time-frequency
dictionary. IEEE Transactions on Signal Processing 41 (1993) 3397.

\bibitem{14}  D. L. Donoho, M. Elad, V. N. Temlyakov, Stable recovery
of sparse overcomplete representations in the presence of noise. IEEE
Transactions on Information Theory, 52 (2006) 618.

\bibitem{15}  M. Andrecut, S. Huang, S. A. Kauffman, 
Heuristic approach to sparse approximation of gene regulatory networks,
Journal of Computational Biology 15(9) (2008) 1173 .

\bibitem{16}  M. Andrecut, S. A. Kauffman, 
On the sparse reconstruction of gene networks, 
Journal of Computational Biology 15(1) (2008) 21 .

\bibitem{17}  NVIDIA, Nvidia CUDA Compute Unified Device Architecture,
Programming Guide, 2008.

\bibitem{18}  NVIDIA, CUBLAS Library, 2008.

\bibitem{19}  M. Andrecut, Fast GPU implementation of sparse signal recovery
from random projections, Engineering Letters (to appear, 2009,
arXiv:0809.1833),
www.nvidia.com/content/cudazone/CUDABrowser/assets/data/applications.xml.
\end{thebibliography}
\end{document}